# Preparation of thin PbTiO$_3$ Nanotubes by the Electrophoretic Deposition Method

Ehsan Rahmati Adarmanabadi[1], Abolghasem Nourmohammadi[2]*, Seyed Mohammad Hassan Feiz[1], and Maryam Lanki[1]

[1] Department of Physics, Faculty of Science, University of Isfahan, 81746-73441, Isfahan, Iran

[2] Department of Nanotechnology, Faculty of Advanced Science and Technologies, University of Isfahan, 81746-73441, Isfahan, Iran

*Email: a.nourmohammadi@sci.ui.ac.ir

Phone: +98-311-7934718

Fax: +98-311-7934404

**Abstract—** Sol-gel electrophoresis is used to grow PbTiO$_3$ nanotube arrays in porous anodic alumina template channels, because it is a cheap and simple method for the growth of nanostructures and has the advantage of better tube growth control. Moreover, this method can produce nanotubes with high quality and more condense structures. In this technique, semiconductor porous anodic alumina templates are used to grow the nanotube arrays. Consequently, close-packed PbTiO$_3$ nanotube arrays are grown in the template channels. It is shown here that, to the best of our knowledge, sol-gel electrophoresis is the only method, applicable for producing PbTiO$_3$ nanotubes with thickness below 20 nm (section 3.3). Also, the effect of deposition time on the wall thickness is investigated, for a fix electrophoresis voltage. The thickness of the grown nanotubes is uniform; an important issue for the ferroelectric properties of the deposited nanolayers for future investigations.

# 1. Introduction

Ferroelectric nanotubes have attracted considerable attention because of their superior properties and potential technological applications such as mass storage memories, energy-harvesting devices, and advanced sensors[1, 2].

Lead titanate or $PbTiO_3$ is a well-known ferroelectric material with the tetragonal perovskite structure. It is a solid solution of PbO and $TiO_2$ and has the highest Curie temperature in the $PbTi_{(x)}Zr_{(1-x)}O_3$ (PZT) family. Nowadays, $PbTiO_3$ nanotubes are very attractive nanomaterials. They are considered for many important applications in nanoelectronics [3, 4]. These nanomaterials have very interesting properties which can make them promising for the next generation ferroelectric nanodevices. Ferroelectric phase transition in most of the perovskite ultrafine layers strongly suggests the presence of a critical size, below which the ferroelectric state becomes unstable [5]. However, recent investigations predict absence of the ferroelectric critical size in ultrathin PbTiO3 nanotubes [6]. Hence, nanotubes of this material are in focus of investigations both theoretically [7] and experimentally [8].

Many attempts have been made in fabrication of $PbTiO_3$ nanotube arrays. Those arrays are already fabricated employing different techniques such as metalorganic chemical vapor deposition (MOCVD), hydrothermal process and sol-gel method[3, 4, 9, 10, 11, 12].

The researchers here have used sol-gel electrophoresis to grow $PbTiO_3$ nanotube arrays in porous anodic alumina template channels, because of low costing and simple method for the growth of nanostructures. This method has already been used for the preparation of $PbTiO_3$ nanotubes by us [8]. It makes a high quality and more condense nanotubes compared with other sol-gel methods. In this study, particularly, it is shown here that sol-gel electrophoresis can be used for producing PbTiO3 nanotubes with thickness below 20 nm. Also, the effect of the anodizing parameters on the diameter of the template pores, and effect of electrophoresis parameters on the wall thickness were investigated. All these effects were confirmed by scanning and transmission electron microscopy investigations.

## 2. Experimental

### 2.1 Preparation of nanoporous alumina templates

In the first section of this study, semiconductor alumina templates were prepared via two-step-anodizing of pure aluminum foils in phosphoric acid. Details of fabricating the alumina templates were already report elsewhere[13]. Briefly, aluminum foils (99.92% purity and 0.3mm thick, Merck) were used to prepare porous anodic alumina templates for growing $PbTiO_3$ nanotubes as deposition templates. First, these aluminum foils were degreased in ethanol and then in acetone. They were annealed at 500°C for 4 hours to increase their grain size. Then, in order to reduce the surface roughness, these foils were electropolished in 2:4:4 volume solution of $H_2O$: $H_2SO_4$: $H_3PO_4$. Anodic alumina templates were produced on an aluminum substrate via a two-step-anodizing process. In the first step, the anodic layer was grown in a 10wt.% solution of phosphoric acid at 2°C under constant voltage of 120V. The grown anodic oxide layer was subsequently removed in a mixture of phosphoric acid (6wt. %) and chromic acid (1.8wt. %) at 60°C. The second anodizing step was carried out under the same anodizing conditions. The anodizing current was measured during both of the first and second anodizing steps. At the end of anodizing, the aluminum substrate was dissolved in a saturated solution of $HgCl_2$ to obtain the transparent alumina templates.

## 2.2 Preparation of the $Pb_{1.03}TiO_3$ sol

The produced porous templates were filled with a highly stabilized lead titanate sol. An extra Pb content 3 mol% was added to the sols in order to improve crystallization. For preparing the $Pb_{1.03}TiO_3$ precursor sol, 19.63gr lead acetate trihydrate (99.5% purity, Merck) was dissolved in 17.43ml pure glacial acetic acid. Then, the prepared solution was dehydrated at 110ºC for 15min and cooled down to room temperature. After cooling down, 17.36ml titanium butoxide (98% purity, Merck) was added gradually drop by drop to the solution. For 15min, this precursor solution was stirred at room temperature, and then, a mixture of 9ml de-ionized water and 24.49ml ethanol (99.5% purity, Merck) was added to cause hydrolysis and to prevent fast gelation of the prepared sol. Then, 2.9ml Ethylene glycol (99% purity, Merck) and 5.19ml acetyl acetone (99% purity, Merck) were added to enhance viscosity and stability of the final solution.

## 2.3 Electrophoretic growth of $PbTiO_3$ nanotubes

In this section, template pores were filled with a highly stabilized $Pb_{1.03}TiO_3$ sol by applying a DC electric voltage. To do this, the template pores were etched with a 5wt. % of phosphoric acid solution for 165min at room temperature. To make the conducting electrodes, the bottom of the etched template was coated with two 100nm conducting layers. For the first conducting layer (bottom layer), the aluminum oxide surface was coated with copper employing the sputtering method. For the second conducting layer (top layer), an aluminum layer was deposited with the same procedure on the top of the bottom copper layer. Afterwards, the templates channels were electrophoretically filled with the prepared $Pb_{1.03}TiO_3$ solution by applying different electrophoresis parameters (1-4V, 3-10). Accordingly, the electrophoretic growth of the nanotubes were observed on the template pore walls.

## 2.4 Heat treatment and characterization

For the crystallization of single tetragonal perovskite $PbTiO_3$ phase, the optimized heat treatment conditions of our previous study, [14], were used. Briefly, the filled templates were dried at 100°C in air ambient for 10 hours and then fired at 680°C in air for 6 hours to crystallize the desired tetragonal perovskite structure.

The prepared nanotubes were characterized using a scanning electron microscope (SEM) system model VEGA2 TESCAN and a transmission electron microscope (TEM) system model Philips EM208S. The phase structure of the nanotubes was analyzed by the X-Ray diffraction (XRD) technique on a Bruker-D8 advance model X-Ray diffractometer using Cu-Kα radiation to show tetragonal structure of prepared nanotubes. The average nanocrystallite size of the grown nanotubes was calculated by the Williamson-Hall method, based on XRD data.

In order to apply the electrophoretic voltage and measure the electrophoretic current simultaneously, a Keithley 2400 LV digital source-meter was employed. An electrophoretic voltage (in range of 1- 4V) was applied and current intensities were measured over different periods of time. In order to observe the grown nanotubes, the filled templates after firing, were polished by corundum nanopowder for 20min in order to remove the top surface layer. Then, the polished alumina templates were etched away in 5wt. % NaOH solution at room temperature for 7 min.

3. **Results and discussion**

**3.1 Porous alumina templates**

SEM images of one alumina template which is produced in this study are shown in Figure 1 (a), (b). The alumina template in Figure 1 was fabricated by two-step-anodizing process under constant voltage of 120V and the electric current was decreased down to ~5mA in a 10wt. % phosphoric acid solution. The first step of anodizing was carried out for 5 hours and, after etching away the oxide layer formed in the first step, the second anodizing step was performed for 18 hours. According to the SEM image of the template, the average pores size is about 180 nm (Figure 1).

**3.2 Effect of the electrophoretic voltage**

In continuation of our previous work[8], the dependence of the thickness of the heat treated nanotubes on the electrophoretic parameters was considered here. Also, the variation of wall thickness versus voltage was studied. The variation of wall thickness of $PbTiO_3$ nanotubes grown at various electrophoretic voltages, for fixed period of 10min, is shown in Figure 2. As observed in this figure, an increase in the electrophoretic voltage leads to a nonlinear increase in the thickness of the prepared nanotubes. Thus, close-packed $PbTiO_3$ nanotubes, with the average thickness of 27-39nm, were produced by the application of DC voltages of 1-4 Volts to the $Pb_{1.03}TiO_3$ precursor sol for 10 min. Thick $PbTiO_3$ nanotubes were obtained at 4 volts (Figure 3). It is found from figure 3 that mechanical polishing by corundum powder (section 2.4) is not an effective and satisfactory method for preparing well-ordered nanotube arrays from the electrophoretically grown $PbTiO_3$ nanotubes. It should be noted that, to prepare well-ordered nanotube arrays, it is preferable to polish the filled template surfaces by more advanced milling techniques such as ion beam milling method, which is not available in our laboratory.

**3.3 Preparation of thin PbTiO$_3$ Nanotubes**

Figure 4 shows the effect of the deposition time on the thickness of the grown nanotubes under the constant voltage of 1V. The thickness of the grown nanotube walls was non-linearly increased with an increase in deposition time during the electrophoretic process. As seen in figure 4, the wall thickness of nanotubes increases rapidly in the initial stages of deposition. But, rate of thickness variation reduces slowly and gradually at longer periods of time. In constant-voltage electrophoresis, the potential between the electrodes is maintained constant, but with increasing deposition (and therefore, increasing the electrical resistance of the deposited layer) the electrophoretic field (E, V/cm) decreases; so does the deposition rate[15]. Thin PbTiO$_3$ nanotubes were finally obtained after 3 minutes of the sol electrophoretic deposition (Figure 5).

It is observed in Figure 5 that PbTiO3 nanotubes with thickness below 20 nm have been successfully grown, at 1 volt, by our low-cost sol-gel electrophoresis deposition method.

**3.4 XRD data**

In preparing XRD patterns due to the prepared nanotubes, we suffered many problems in our previous work, [8], because alumina templates are amorphous and produce a great deal of XRD background noise in the diffraction pattern. For this reason, identifying of the phase structure of PbTiO$_3$ via XRD was very difficult. Therefore, more studies were carried out on this issue. Typical XRD patterns of the prepared

PbTiO$_3$ nanotubes are shown in Figures 6a and 6b. The template of figures 6a was filled electrophoretically by the PbTiO$_3$ sol at the fixed voltage of 4V for 10 min. The filled template was dried at 100°C for 10 hours and fired at 680°C for 6 hours. As observed in this figure, the tetragonal perovskite PbTiO$_3$ structure is crystallized as the main phase according to the JCPDS card number 6-452.

It is clearly seen that by using Pb$_{1.03}$TiO$_3$ stoichiometry, highly crystalline and single phase lead titanate nanotubes were produced at 680°C. The average nanocrystallite size of the grown nanotubes was calculated from the Williamson-Hall plot (not shown here) came out to be about 7 nm.

### 3.5 TEM Results

In complementary of SEM and XRD studies, TEM structural and electron diffraction investigations were performed. Figure 7 shows the TEM micrograph of a PbTiO$_3$ nanotube. To observe this nanotube, first, surface of the filled alumina template was polished using corundum nanopowder. Then, the alumina template was heavily etched away in a 5%.wt caustic soda solution. Consequently, free-standing PbTiO$_3$ nanotubes were produced. These nanotube were subsequently dispersed in pure ethanol by an ultrasonic shaker.

The uniformity of tube wall thickness is clearly observed in Figure 7. This is due to the effect of the applied electric field which makes uniformity and thickness controllable.

The electron diffraction pattern of the analyzed nanotube is shown in Figure 8. This figure confirms the existence of single-phase tetragonal perovskite PbTiO$_3$ nanotubes after heat treatment at 680°C.

**3.6 Advantages of our research work**

In the last decade, many efforts are made in order to reach high quality, single-phase and highly crystalline ferroelectric nanotubes, through the sol-gel based methods[16]. However, it is well known that in the conventional sol-gel process, controlling the nanotube thickness is a difficult problem. It should be noticed that the stability of the ferroelectric phase in ferroelectric materials strongly depend on the grain size, and on the other hand, it relates to the thickness of the deposited layer [17, 18].

Per Martin Rørvik and coworkers,[3], have fabricated PbTiO3 nanotubes via the conventional template-assisted sol-gel synthesis method. Their nanotubes had various outer diameters of 200–400 nm due to application of various alumina templates. But, all their nanotubes wall thickness was about 20 nm, as reported by them. Here, the wall thickness of grown nanotubes was altered, in a wide range, by changing the electrophoretic parameters. Also, various anodizing voltage were used by us to control the template pore diameters in the range of 80-200 nm in order to grow nanotubes with different outer diameters.

To the best of our knowledge, among the reserchers, only Fujisawa and coworkers have achieved $PbTiO_3$ nanotubes with a wide range of thickness (40-100nm). Fujisawa and coworkers have used the MOCVD method for preparation of PbTiO3 nanotubes [13]. MOCVD is a well known technique to fabricate high-quality thin films. However, this method is very expensive and time consuming and various accessories are required to deposit nanotubes of $PbTiO_3$ material via this method. On the other hand, our sol-gel electrophoresis method is a simple and cost-effective method to grow nanotubes with different wall thickness.

Macak and coworkers produced PbTiO$_3$ nanotubes using the hydrothermal method. In this method, the titanium foils are anodized first to produce TiO$_2$ nanotubes, and, then, the solid Pb is defused into the TiO$_2$ structure by electrical deposition method. The thickness of nanotubes which they have prepared were in the range of 20-32nm [10]. But, they were composed of several different materials such as PbTiO$_3$, PbO, as well as both TiO$_2$ rutile and anatase polymorphs.

Here, to consider our nanotubes for future investigations, first, we investigated the controllability of the thickness of the deposited layers. After firing, single tetragonal phase PbTiO$_3$ nanotubes, with the average thickness of 27-39nm were produced here by the application of DC voltages of 1-4 Volts to the Pb$_{1.03}$TiO$_3$ precursor sol for 10 min. In addition, thin PbTiO$_3$ nanotubes of about 17-27nm were produced by the application of DC voltages of 1 Volt to the Pb$_{1.03}$TiO$_3$ precursor sol for 3-10 min.

In our previous work [8], we did not consider the TEM investigation. Thus, we were not able to confirm uniform growth of nanotube walls. In the current research work we studied more precisely than our previous one, using TEM data as a complement of our work, because only TEM is capable to show the uniformity of wall thickness of nanotubes. Our TEM micrographs confirmed that PbTiO$_3$ nanotubes with uniform thickness are produced.

## 4. Conclusions

In the conventional sol-gel process, controlling the nanotube thickness is a difficult problem. In this article, sol-gel electrophoresis process was successfully utilized as a new way to fabricate perovskite PbTiO$_3$ nanotubes with controlled wall thickness. Besides, this method is cheap and simple, and by using this way, a high quality and more condense nanotube were obtained. The growth of nanotubes was controlled by electrophoresis parameters (the applied voltage and deposition time). Also, the uniformity of tube wall thickness was confirmed by TEM studies. In our method, it was possible to reach single tetragonal phase $PbTiO_3$ nanotubes. Extra Pb content was added to the precursor sol in order to improve crystallization of the grown amorphous $PbTiO_3$ nanotubes. Highly crystalline and single tetragonal phase $PbTiO_3$ nanotubes with the average thickness of 27-39nm were produced by the application of DC voltages of 1-4 Volts to the $Pb_{1.03}TiO_3$ precursor sol for 10 min. Moreover, to the best of our knowledge, $PbTiO_3$ nanotubes, with low thickness of about 17-27nm, were produced by our method, for the first time, by the application of DC voltages of 1 Volt to the sol for 3-10 min. In addition, by using various anodizing voltages, alumina template with different pore diameters in the range of 80-200 nm were produced in order to grow nanotubes with different outer diameters.

**Acknowledgment**

The authors would like to thank the Office of Graduate Studies of the University of Isfahan for their support.

**Figure Captions**

**Fig1.** SEM image of (a) top surface and (b) cross section of a porous anodic alumina template produced via two-step anodizing at 120V.

**Figure 2.** Effect of the applied voltage: average thickness of the PbTiO$_3$ nanotubes grown in different voltages for 10min.

**Figure 3**. SEM image of PbTiO$_3$ nanotubes fabricated at the electrophoretic voltage of 4V for 10min.

**Figure 4.** Effect of the deposition time: Average wall thickness of nanotubes grown at the fixed voltage of 1 V, during 0-10 min of deposition.

**Figure 5.** SEM images of PbTiO$_3$ nanotubes fabricated at 1 volt for (a) 3 min, (b) 10 min.

**Figure 6.** XRD diffraction patterns of the alumina template filled with lead titanate nanotubes grown under different conditions; (a) 4V for 10min, (b) 2V for 15min.

**Figure 7.** The TEM pattern of prepared PbTiO$_3$ nanotubes heat treated at 680°C for 6 h.

**Figure 8.** The electron diffraction pattern of the PbTiO$_3$ nanotubes in Figure 6.

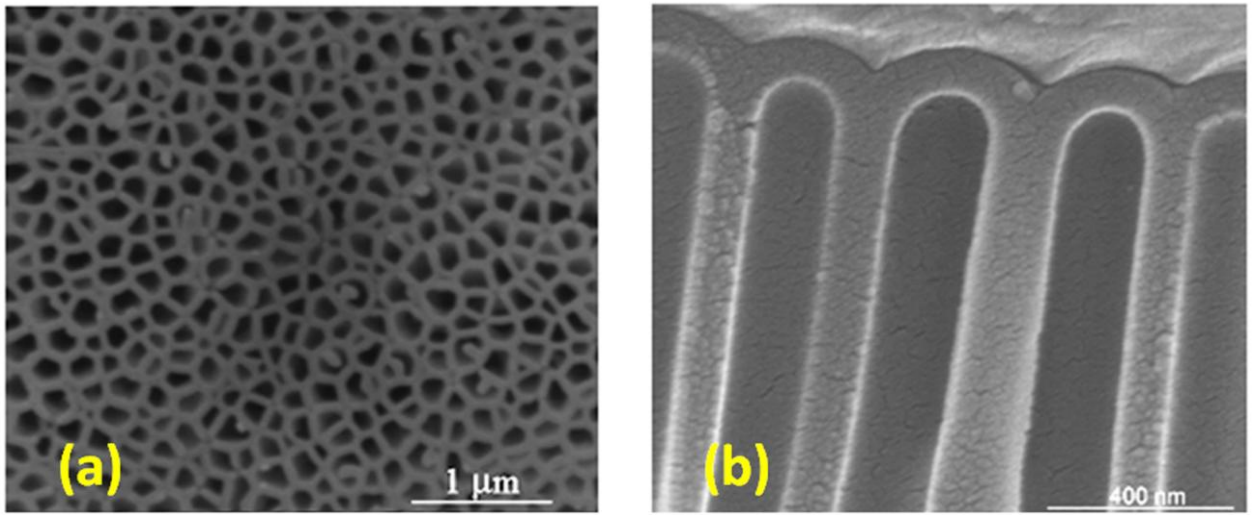

Figure 1

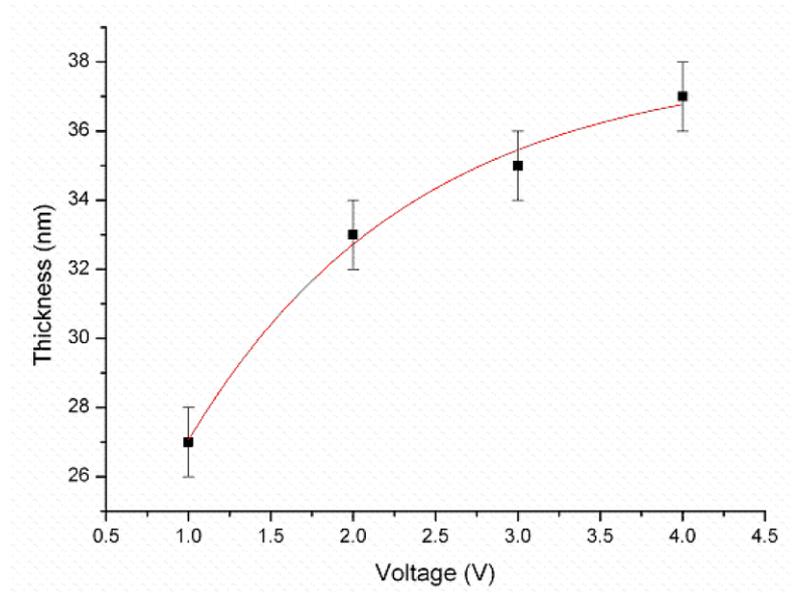

Figure 2

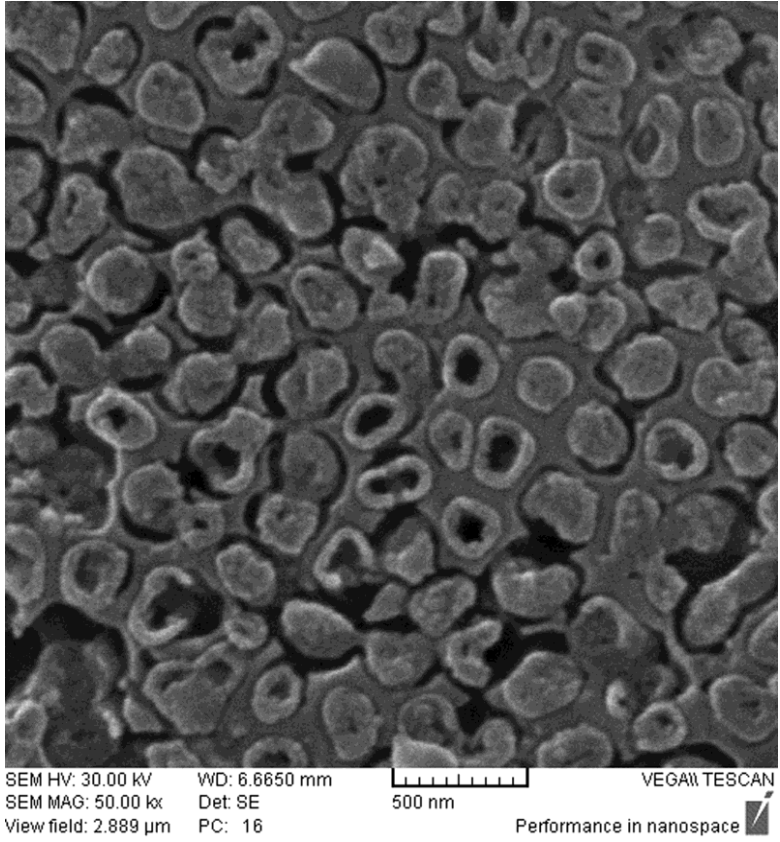

Figure 3

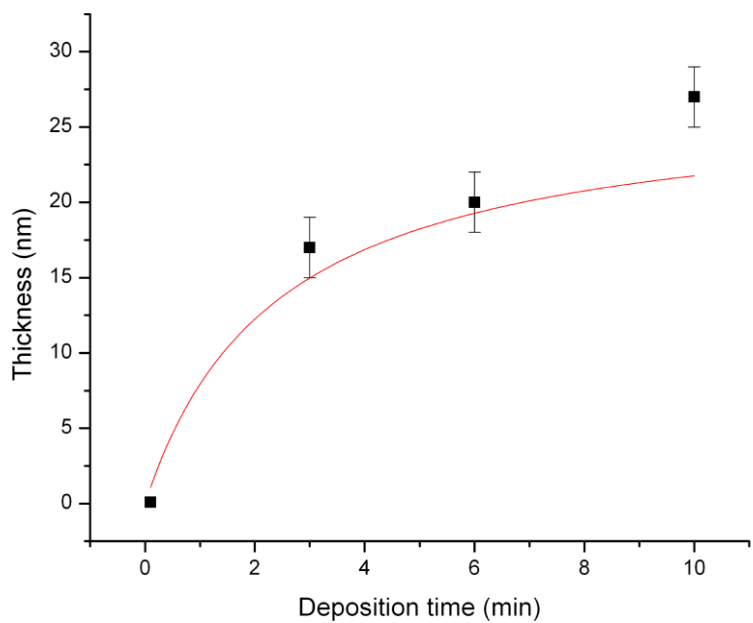

Figure 4

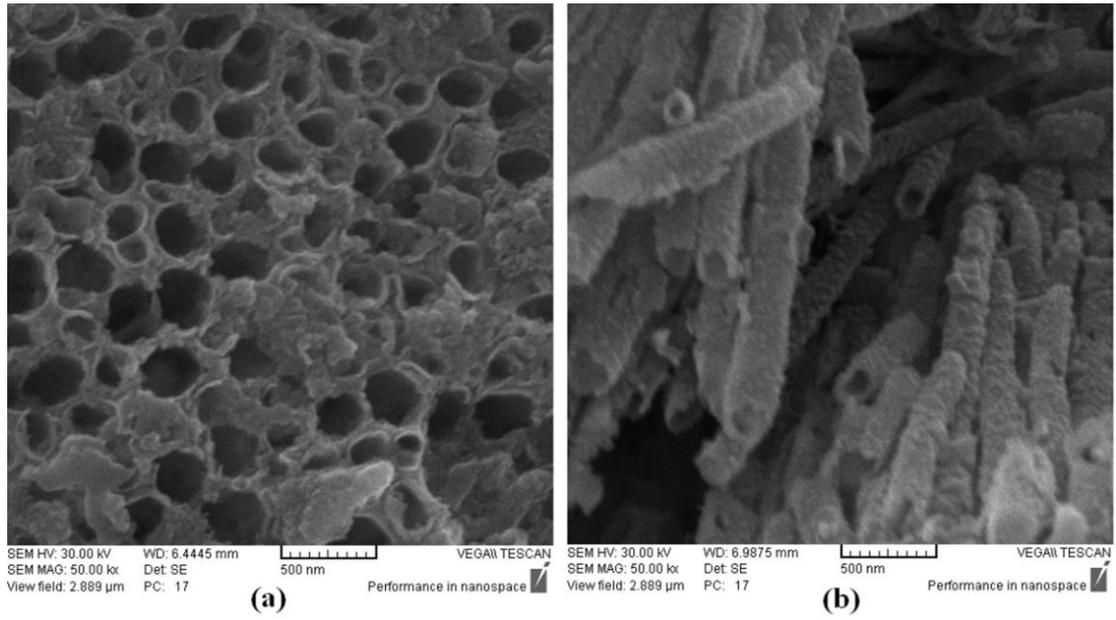

Figure 5

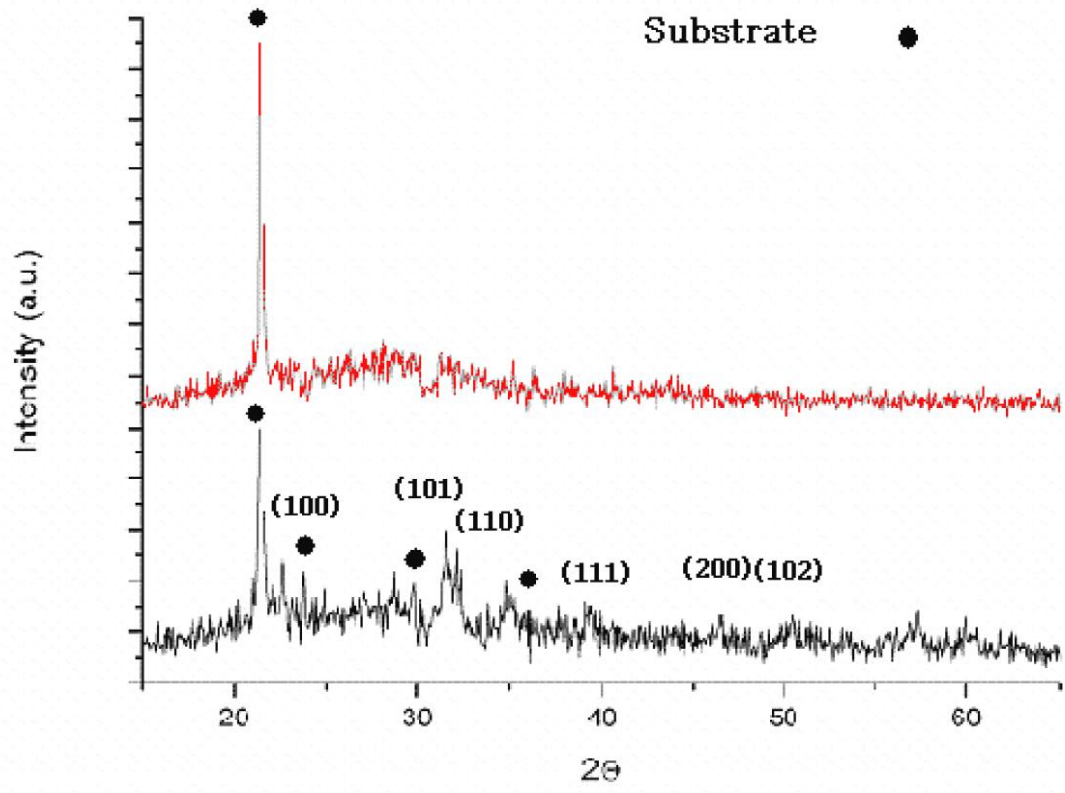

Figure 6

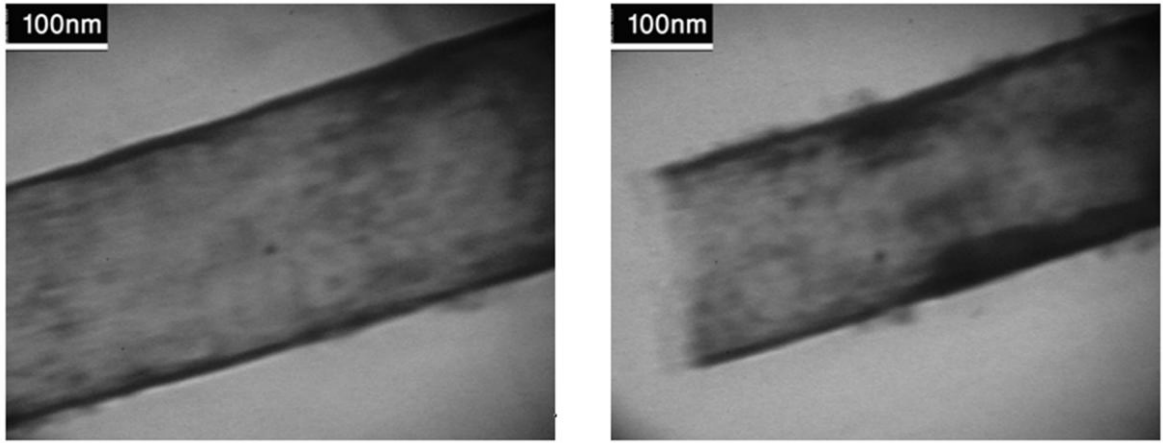

Figure 7

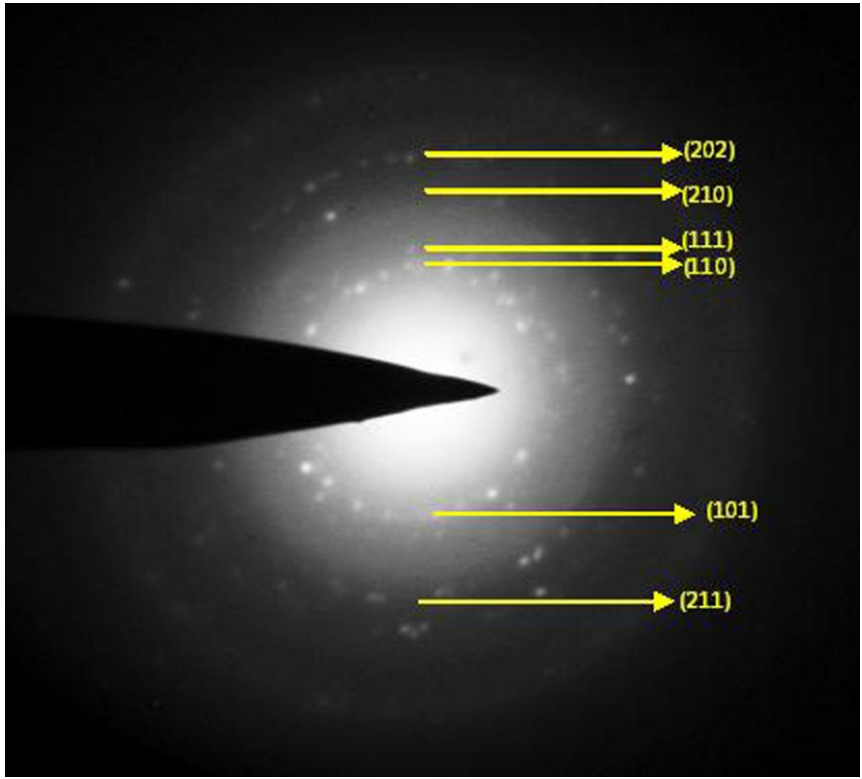

Figure 8